\documentclass[twoside,slac_one]{revtex4}
\usepackage{graphicx}
\usepackage{fancyhdr}
\usepackage{amsmath} 
\usepackage{bm}
\usepackage{amsxtra}
\usepackage{amssymb}
\usepackage{amsthm}
\usepackage{latexsym}
\usepackage{lscape}

\begin{document}

\title{Chiral field theory of $0^{-+}$ glueball}

%

\author{Bing An Li}
\affiliation{Department of Physics and Astronomy, University of Kentucky,USA}

\begin{abstract}
A chiral field theory of $0^{-+}$ glueball is presented. The coupling between the quark operator and the $0^{-+}$ glueball
field is revealed from the U(1) anomaly. 
The Lagrangian of this theory is constructed
by adding a $0^{-+}$ glueball field to a
successful Lagrangian of chiral field theory of pseudoscalar, vector, and axial-vector mesons.
Quantitative study of the physical processes of the $0^{-+}$ glueball of $m=1.405\textrm{GeV}$ is presented.
The theoretical predictions can be used to identify the $0^{-+}$ glueball.
\end{abstract}

\maketitle

\thispagestyle{fancy}


\section{Introduction}
Glueball is the solution of nonperturbative QCD and there is extensive study
on pseudoscalar glueballs.
On the other hand, many candidates of
$0^{++},\;0^{-+}$, and $2^{++}$ glueballs have been discovered. However, identification of a glueball
is still in question. 
It is the attempt of this talk to present a chiral field theory ~\cite{1} which can do systematic and
quantitative study of the
physical processes of the $0^{-+}$ glueball.

Both current algebra and Lattice QCD successfully use quark operators to study nonperturbative hadron physics.
Based on current algebra and QCD
we have proposed a chiral field theory of pseudoscalar, vector, and axial-vector mesons ~\cite{2}. 
The Lagrangian of quarks and mesons is constructed as
\begin{eqnarray}
{\cal L}_{1}=\bar{\psi}(x)(i\gamma\cdot\partial
+\gamma\cdot v+\gamma\cdot a\gamma_{5}
-mu(x))\psi(x)-\bar{\psi}M\psi\nonumber \\
+{1\over 2}m^{2}_{0}(\rho^{\mu}_{i}\rho_{\mu i}+
\omega^{\mu}\omega_{\mu}+a^{\mu}_{i}a_{\mu i}+f^{\mu}f_{\mu}
+K^{a}_{\mu}\bar{K}^
{a\mu}+K_{1}^{\mu}K_{1\mu}
+\phi_{\mu}\phi^{\mu}+f_{s}^{\mu}f_{s\mu})
\end{eqnarray}
where \(a_{\mu}=\tau_{i}a^{i}_{\mu}+\lambda_{a}K^{a}_{1\mu}
+({2\over 3}+{1\over \sqrt{3}}\lambda_{8})
f_{\mu}+({1\over 3}-{1\over \sqrt{3}}\lambda_{8})
f_{s\mu}\)(\(i=1,2,3\) and \(a=4,5,6,7\)),
\(v_{\mu}=\tau_{i}
\rho^{i}_{\mu}+\lambda_{a}K^{a}_{\mu}+
({2\over 3}+{1\over \sqrt{3}}\lambda_{8})
\omega_{\mu}+({1\over 3}-{1\over \sqrt{3}}\lambda_{8})
\phi_{\mu}\),
\(u=exp\{i\gamma_{5}
(\tau_{i}\pi_{i}+
\lambda_{a}K^{a}+\lambda_8\eta_8+
{1\over\sqrt{3}}\eta_0)\}\),
\it{m} is the constituent quark mass which originates in the quark condensate,
\it{M} is the matrix of the current quark mass, $m_0$ is a parameter.
In the limit, $m_q\rightarrow 0$, the theory (1) has global $U(3)_L\times U(3)_R$ symmetry.
In this theory the meson fields are related to corresponding quark operators. For example,
at the tree level the vector and the axial-vector mesons are expressed as the quark
operators
\[\rho^i_\mu=-{1\over m^2_0}\bar{\psi}\gamma_\mu\tau^i\psi,\;\;
a^i_\mu=-{1\over m^2_0}\bar{\psi}\gamma_\mu\gamma_5\tau^i\psi.\]
The pseudoscalar mesons are via the mechanism of the nonlinear $\sigma$ model introduced. This mechanism makes this 
theory different from the NJL model.
In QCD the mesons are not independent degrees of freedoms.
Therefore, there are no kinetic terms for meson fields in Eq. (1). Only one quark loop calculation is allowed from Eq. (1).
Integrating out the quark fields,
the Lagrangian of the meson fields is derived. This procedure is equivalent to one quark loop calculation. 
The process deriving the Lagrangian of mesons has nonperturbative nature.
The kinetic terms of the meson fields are generated by the quark loop diagrams.
$N_c$ expansion is naturally embedded. The tree diagrams are at the leading
order and the loop diagrams of the mesons are at the higher orders.
The major features of nonperturbative QCD: $N_C$ expansion, quark condensate, and chiral symmetry are all included in
this meson theory.

In most cases there are two parameters, $f_\pi$ and a universal coupling constant g which is determined by the decay rate of
$\rho\rightarrow ee^+$ to be 0.395. This meson theory is phenomenologically successful under 2$\textrm{GeV}$.
The masses of the mesons are determined. The pseudoscalar mesons are Gladstone bosons and in the chiral limit
$m_q\rightarrow 0$ the masses of pseudoscalar mesons vanish. 
\[m^2_\rho = 6 m^2.\]
The masses of the vector and the axial-vector mesons are the same in the L (1). The fact is that the $a_1(1260)$ is much heavier than
the $\rho(770)$ meson. The Weinberg's second sum rule tells \(m^2_a = 2 m^2_\rho \).    
A complete new mechanism of generating mass for axial-vector mesons is revealed from Eq. (1). The kinetic term, gauge fixing,
and additional mass of the $a_1$ field are revealed from the vacuum polarization of the $a_1$ field
\begin{equation}
(1-\frac{1}{2\pi^2 g^2})m^2_{a_1} = 2 m^2_\rho.
\end{equation}
Comparing with the Weinberg's second sum rule, there is an additional factor. An assumption has been made in deriving 
the Weinberg's second sum rule. Getting rid this assumption, the additional factor is obtained. By the way, this new mass formula
fits the data much better.

This new mechanism happens to both nonconserved axial-vector current and nonconserved charged vector currents.
Using this new mechanism, a new Electro-Weak theory without spontaneous symmetry breaking and Higgs has been proposed ~\cite{3}.
The Lagrangian of this new EW theory with $SU(2)_L\times U(1)$ gauge fields is constructed as
\begin{eqnarray}
\lefteqn{{\cal L}=
-{1\over4}A^{i}_{\mu\nu}A^{i\mu\nu}-{1\over4}B_{\mu\nu}B^{\mu\nu}
+\bar{q}\{i\gamma\cdot\partial-M\}q}
\nonumber \\
&&+\bar{q}_{L}\{{g\over2}\tau_{i}
\gamma\cdot A^{i}+g'{Y\over2}\gamma\cdot B\}
q_{L}+\bar{q}_{R}g'{Y\over2}\gamma\cdot Bq_{R}\nonumber \\
&&+\bar{l}\{i\gamma\cdot\partial-M_{f}\}l
+\bar{l}_{L}\{{g\over2}
\tau_{i}\gamma\cdot A^{i}-{g'\over2}\gamma\cdot B\}
l_{L}-\bar{l}_{R}g'\gamma\cdot B l_{R}.
\end{eqnarray}
The $SU(2)_L\times U(1)$ symmetry is broken by the fermion masses. There are no mass terms for W and Z bosons.
Therefore, the perturbation theory involving internal lines of W and Z cannot be defined.
However, one-loop fermion diagrams, only one-loop, are calculable. In The L (3) there are both the axial-vector
and the charged vector currents and massive fermions. The new mechanism mentioned above can be applied.
The vacuum polarization of the Z field is expressed as
\[\Pi_{\mu\nu}(q^2)= F_1(q^2)(q_\mu q_\nu-q^2 g_{\mu\nu})+F_2(q^2)q_\mu q_\nu+{1\over2}\Delta m^2_Z g_{\mu\nu}.\]
Therefore, both the gauge fixing term($F_2$) and the mass term of the field are dynamically generated from
the fermion masses. Top quark mass plays a dominant role.
Non zero $\partial_\mu Z^\mu$ leads to a scalar field and a gauge fixing term for the Z field. 
The mass of the scalar field is determined to be
\[m_{\phi^0}=m_t e^{\frac{m^2_Z}{m^2_t}\frac{16\pi^2}{3\bar{g}^2}+1}=3.78\times10^{14}GeV.\]
The gauge fixing is calculated to be
\[\xi_z=-1.18\times10^{-25}.\]
After renormalization of the mass term of the Z boson it is obtained
\[m_z={1\over2}\bar{g}^2 m^2_t.\]
It agrees well with the data.
Similarly, the vacuum polarization of the W boson is calculated. A charge scalar field is dynamically generated
\[m_{\phi^{\pm}}=m_t e^{\frac{m^2_W}{m^2_t}\frac{16\pi^2}{3g^2}}=9.31\times10^{13}GeV.\]
The gauge fixing is determined
\[\xi_W=-3.73\times10^{-25}.\]
After renormalization the mass of the W boson is determined as
\[m^2_W={1\over2}g^2 m^2_t,\;\;G_F=\frac{1}{2\sqrt{2}m^2_t}.\]
It agrees well with the data. It also obtain
\[\frac{m^2_W}{m^2_Z}=\frac{g^2}{\bar{g}^2}=cos^2\theta_W.\]
The propergators of Z- and W- fields are derived as
\[\Delta_{\mu\nu}^Z=
\frac{1}{q^2-m^2_Z}\{-g_{\mu\nu}+(1+\frac{1}{2\xi_Z})\frac{q_\mu q_\nu}{
q^2-m^2_{\phi^0}}\},\;\;
\Delta^W_{\mu\nu}=
\frac{1}{q^2-m^2_W}\{-g_{\mu\nu}+(1+\frac{1}{2\xi_W})\frac{q_\mu q_\nu}{
q^2-m^2_{\phi_W}}\}.\]
In this theory there is no quadratic divergence. The neutral scalar is always associated with the factor 
${1\over m_Z}$ and the charged scalar is always associated with the factor ${1\over m_W}$. These factors make the 
interactions of the scalars with other matter fields weaker.
As an example, the interaction of the charged scalar with EM fields is found to be 
\[{\cal L}=16\sqrt{2}i{e\over g^2}G_F F^{\mu\nu}\partial_\mu\phi^+\partial_\nu\phi^- .\]
The strength of the electromagnetic interactions of the charged scalars is at the order of weak interactions.

The EM fields can be added into the L (1). the Vector Meson Dominance(VMD) is a natural result of the L. The ChPT is the
low energy limit of this theory and all 10 coefficients of the ChPT are predicted.
The triangle and the WZW anomaly are derived from the imaginary part of the L (1). The decay widths of strong, 
EM, and weak interactions of the mesons are calculated. New results for pion and kaon form factors
are obtained. The $\pi-\pi$ and $\pi-K$ scattering are calculated.
Theory agrees with data very well.

\section{Chiral Lagrangian of $0^{-+}$ glueball and mesons}

The Lattice Gauge Theory has used the gluon operator, $F\tilde{F}$(in the continuum limit),
to calculate the mass of the pseudoscalar glueball by the quench approximation.
Under the least coupling principle the effective Lagrangian is constructed as
\begin{eqnarray}
{\cal L}=-{1\over4}F^{a\mu\nu}F^{a}_{\mu\nu}+F^{a}_{\mu\nu}\tilde{F}^{a\mu\nu}\chi+{1\over2}G^2_\chi \chi\chi,\;\;
\chi=-{1\over G^2_\chi}F_{\mu\nu}\tilde{F}^{\mu\nu}.
\end{eqnarray}
In QCD glueball is a bound state of gluons, not an independent degree of freedom, therefore,
there is no kinetic term for the glueball field $\chi$.
The relationship between the gluon operator $F\tilde{F}$ and the quark operators is found from the U(1) anomaly
\begin{equation}
\partial_\mu(\bar{\psi}\gamma_\mu\gamma_5\psi)=2i\bar{\psi}M\gamma_5\psi+\frac{3g^2_s}{(4\pi)^2}F_{\mu\nu}\tilde{F}^{\mu\nu}.
\end{equation}
The coupling between the glueball field $\chi$ and the quark operator is found
\begin{equation}
{\cal L}=-\bar{\psi}\gamma_\mu\gamma_5\psi\partial_\mu\chi
+{1\over2}G^2_\chi \chi\chi.
\end{equation}
Adding this L to the L of the mesons (1), the L of mesons and $0^{-+}$ glueball is constructed. Integrating out the quark fields
the L of the mesons and the $0^{-+}$ glueball is obtained.
The normalized $\chi$ field is determined as
\begin{equation}
\chi\rightarrow\sqrt{2\over3}{1\over F}\chi.
\end{equation}
F is defined in Ref. ~\cite{2}.
\section{Mass mixing of the $0^{-+}$ glueball $\eta(1405)$ and the $\eta,\; \eta'$}
Phenomenological analysis shows that the $\eta(1405)$ is a possible $0^{-+}$ glueball. Use the constructed L we can test 
whether the $\eta(1405)$ is indeed a $0^{-+}$ glueball.
It is well known that the $\eta'$ contains substantial gluon component.
The mixing of $\eta,\;\eta'\;\eta(1405)$ is studied in this section. Three of the 6 elements of the $3\times 3$ mass matrix
are determined
\begin{eqnarray}
m^2_{\eta_8}=-{4\over f^2_\pi}{1\over3}\langle0|\bar{\psi}\psi|0\rangle{1\over3}(m_u+m_d+4m_s)
={1\over3}\{2(m^2_{K^+}+m^2_{K^0})-m^2_\pi\},\nonumber\\
\Delta m^2_{\eta_8\eta_0}={4\sqrt{2}\over9}{1\over f^2_\pi}{1\over3}\langle0|\bar{\psi}\psi|0\rangle(m_u+m_d-2m_s)
={\sqrt{2}\over9}(m^2_{K^+}+m^2_{K^0}
-2m^2_\pi),\nonumber \\
\Delta m^2_{\chi\eta_8}=-{4\sqrt{2}\over9}{1\over f_\pi F}{1\over3}\langle0|\bar{\psi}\psi|0\rangle (m_u+m_d-2m_s)
=-{\sqrt{2}\over9}{f_\pi\over F}(m^2_{K^+}+m^2_{K^0}-2m^2_\pi).
\end{eqnarray}
Because of the U(1) anomaly $m^2_{\eta_0}$ is taken as a parameter. The physical masses, $m^2_{\eta},\;m^2_{\eta'}
,\;m^2_{\eta(1405)}$ and $\Gamma(\eta'\rightarrow\gamma\gamma)$ are taken as inputs
\begin{equation}
\Gamma(\eta'\rightarrow\gamma\gamma)=\frac{\alpha^2}{16\pi^3}\frac{m^3_{\eta'}}{f^2_\pi}(2\sqrt{2\over3}b_2
+{1\over\sqrt{3}}a_2)^2,
\end{equation}
where $a_2$ and $b_2$ are the two coefficients of $\eta'=a_2\eta_8 + b_2 \eta_0 + c_2 g$.
$f_\pi=0.182\;\textrm{GeV}$ is taken. The experimental data of
$\Gamma(\eta'\rightarrow\gamma\gamma)$ is $4.31(1\pm0.13)\;\textrm{keV}$.

The mixing is determined
\begin{eqnarray}
\eta=0.9742\eta_8+0.1593\eta_0-0.16\chi,\nonumber \\
\eta'=-0.1513\eta_8+0.8208\eta_0-0.551\chi,\nonumber \\
\eta(1405)=-0.003522\eta_8+0.5724\eta_0+0.8199\chi.
\end{eqnarray}

\section{$\eta(1405)\rightarrow\gamma\gamma$ decay}
The couplings between $\eta_8,\;\eta_0$
and $\rho\rho,\;\omega\omega,\;\phi\phi$ are presented in Ref. ~\cite{2}
\begin{eqnarray}
{\cal L}_{\eta_8 vv}=\frac{N_C}{(4\pi)^2}\frac{8}{\sqrt{3}g^2 f_\pi}\eta_8\epsilon^{\mu\nu\alpha\beta}
\{\partial_\mu\rho^i_\nu\partial_\alpha\rho^i_\beta
+\partial_\mu\omega_\nu\partial_\alpha\omega_\beta-2\partial_\mu\phi_\nu\partial_\alpha\phi_\beta\},\nonumber\\
{\cal L}_{\eta_0 vv}=\frac{N_C}{(4\pi)^2}\frac{8\sqrt{2}}{\sqrt{3}g^2 f_\pi}\eta_0\epsilon^{\mu\nu\alpha\beta}
\{\partial_\mu\rho^i_\nu\partial_\alpha\rho^i_\beta
+\partial_\mu\omega_\nu\partial_\alpha\omega_\beta+\partial_\mu\phi_\nu\partial_\alpha\phi_\beta\}
\end{eqnarray}
The VMD leads to following relationships
\begin{eqnarray}
\rho^0_\mu\rightarrow{1\over2}egA_\mu,\;
\omega_\mu\rightarrow{1\over6}egA_\mu,\;
\phi_\mu\rightarrow{-1\over3\sqrt{2}}egA_\mu.
\end{eqnarray}
The couplings
\begin{eqnarray}
{\cal L}_{\eta_8\gamma\gamma}=\frac{\alpha N_C}{4\pi}\frac{8}{\sqrt{3} f_\pi}\eta_8{1\over6}
\epsilon^{\mu\nu\alpha\beta}\partial_\mu A_\nu\partial_\alpha A_\beta,\nonumber\\
{\cal L}_{\eta_0\gamma\gamma}=\frac{\alpha N_C}{4\pi}\frac{8\sqrt{2}}{\sqrt{3} f_\pi}\eta_0{1\over3}
\epsilon^{\mu\nu\alpha\beta}\partial_\mu A_\nu\partial_\alpha A_\beta
\end{eqnarray}
are found.

Only the $\eta_0$ component of $\eta(1405)$ contributes to the $\eta(1405)\rightarrow\gamma\gamma$ decay and
the glueball component $\chi$ is suppressed
\begin{equation}
\Gamma(\eta(1405)\rightarrow\gamma\gamma)=\frac{\alpha^2}{16\pi^3}\frac{m^3_{\eta(1405)}}{f^2_\pi}(2\sqrt{2\over3}b_3)^2
=4.55\;\textrm{keV}
\end{equation}
is predicted. 
The total width of the $\eta(1405)$ is $51.5 \pm 3.4\; \textrm{MeV}$ and
\[B(\eta(1405)\rightarrow2\gamma)=0.87 (1 \pm 0.07)\times10^{-4}\]
is obtained. This small branching ratio is consistent
with that $\eta(1405)$ has not been discovered in two photon collisions.
 
\section{$\eta(1405)\rightarrow\gamma\rho,\;\gamma\omega,\;\gamma\phi$ decays}
The $\eta_8$ component of $\eta(1405)$ is ignored and the $\chi$ component doesn't contribute to the coupling 
of the $\eta(1405)vv$.
The vertex of $\eta(1405)vv$ is determined by the quark component $\eta_0$ only. Using the VMD,
\begin{eqnarray}
{\cal L}_{\eta(1405)\rho\gamma}=\frac{eN_C}{(4\pi)^2}\frac{8\sqrt{2}}{\sqrt{3}f_\pi}b_3\eta_0
\epsilon^{\mu\nu\alpha\beta}\partial_\mu\rho_\nu\partial\alpha A_\beta,\nonumber \\
{\cal L}_{\eta(1405)\omega\gamma}=\frac{eN_C}{(4\pi)^2}\frac{8\sqrt{2}}{3\sqrt{3}f_\pi}b_3\eta_0
\epsilon^{\mu\nu\alpha\beta}\partial_\mu\omega_\nu\partial\alpha A_\beta,\nonumber \\
{\cal L}_{\eta(1405)\phi\gamma}=\frac{eN_C}{(4\pi)^2}\frac{16}{\sqrt{3}f_\pi}b_3\eta_0
\epsilon^{\mu\nu\alpha\beta}\partial_\mu\phi_\nu\partial\alpha A_\beta
\end{eqnarray}
are obtained, where \(b_3=0.5724\).
The numerate results are
\begin{equation}
\Gamma(\eta(1405)\rightarrow\rho\gamma)=0.84 \textrm{MeV},\;\Gamma(\eta(1405)\rightarrow\omega\gamma)=90.3\textrm{kev},
\;\Gamma(\eta(1405)\rightarrow\phi\gamma)=58.2\textrm{kev}.
\end{equation}
The branching ratios of these three decay modes are
\[1.63\times10^{-2} (1 \pm 0.07),\;1.75\times10^{-3}(1 \pm 0.07),\;1.13\times10^{-3} (1 \pm 0.07)\]
respectively.

\section{Kinetic mixing of $\chi$ and $\eta_0$ state}
Besides mass mixing between the pseudoscalar mesons and the $0^{-+}$ glueball
there is kinetic mixing between the $\eta_0$ meson and the $\chi$ glueball.
While the mass matrix is diagonalized and the physical states are determined, however, the matrix of
the kinetic terms might not be diagonalized by these new physical states. The $\rho-\omega$ system is a good example.
Eq. (1) shows that the mass matrix
of the $\rho$ and the $\omega$ mesons is diagonalized. The kinetic terms of the $\rho$ and the $\omega$ fields
are generated by the quark loop diagrams. 
The mixing between the kinetic terms of the $\rho^0$ and the $\omega$ fields is dynamically generated by the quark loops,
which is determined by the mass difference of the current quark masses, $m_d - m_u$, and the electromagnetic interactions. 
\[{\cal L}_{\rho-\omega}=\{-\frac{1}{4\pi^2g^2}{1\over m}(m_d-m_u)
+{1\over24}e^2 g^2\}(\partial_\mu\rho_\nu-\partial_\nu\rho_\mu)
(\partial_\mu\omega_\nu-\partial_\nu\omega_\mu).\]
The coefficient of the kinetic mixing of $\eta_0$ and $\eta(1405)$ is determined by three vertices
\begin{equation}
-\sqrt{{2\over3}}{1\over F}\bar{\psi}\gamma_\mu\gamma_5\psi\partial_\mu\chi - {1\over\sqrt{3}}{c\over g}{2\sqrt{2}\over f_\pi}
\bar{\psi}\gamma_\mu\gamma_5\psi\partial_\mu\eta_0
-im{1\over\sqrt{3}}{2\sqrt{2}\over f_\pi}\bar{\psi}\gamma_5\psi\eta_0.
\end{equation}
By calculating the S-matrix element $\langle\eta_0|S|\chi\rangle $, in the chiral limit the kinetic mixing is found to be
\begin{equation}
-(1-{2c\over g})^{{1\over2}}\partial_\mu\eta_0\partial_\mu\chi.
\end{equation}
This kinetic mixing cannot be referred to the mass mixing.

\section{$J/\psi\rightarrow\gamma \eta(1405)$ decay}
In pQCD the $J/\psi$ radiative decay is described as $J/\psi\rightarrow\gamma gg,\;gg\rightarrow meson$.
Therefore, if the meson is strongly coupled
to two gluons it should be produced in $J/\psi$ radiative decay copiously.
Both the $\eta'$ and the $\eta(1405)$ contain large components of the pure glueball state $\chi$. Therefore,
large branching ratios of $J/\psi\rightarrow\gamma\eta',\;
\gamma \eta(1405)$ should be expected.

The decay width of the $J/\psi\rightarrow\gamma\chi$ is derived as ~\cite{4}
\begin{equation}
\Gamma(J/\psi\rightarrow\gamma\chi)=\frac{2^{11}}{81}\alpha
\alpha^2_s(m_c)\psi^2_J(0)f^2_G{1\over m^8_c}\frac{(1-{m^2\over m^2_J})^3}{\{1-2\frac{m^2}{m^2_J}
+{4m^2_c\over m^2_J}\}^2} \nonumber \\
\{2m^2_J-3m^2(1+{2m_c\over m_J})-16{m^3_c\over m_J}\}^2,
\end{equation}
where $\psi_J(0)$ is the wave function of the $J/\psi$ at the origin, $f_G$ is a parameter
related to the glueball state $\chi$, m is the mass of the physical state which contains the $\chi$
state and it will be specified.
After replacing corresponding quantities, $m_c\rightarrow m_b$, $m_J\rightarrow m_\Upsilon$,
$Q_c={2\over3}\rightarrow Q_b=-{1\over3}$, the equation has been applied to study $B(\Upsilon(1S)\rightarrow\gamma
\eta'(\eta))$ and very strong suppression by the mass of the b quark has been found in these processes.
The suppression leads to very small $B(\Upsilon(1S)\rightarrow\gamma
\eta'(\eta))$, which are consistent with the experimental upper limits of $B(\Upsilon(1S)\rightarrow\gamma
\eta'(\eta))$.

The $\chi$ state is via both the mass mixing and the kinetic mixing related to the $\eta'$ and
the $\eta(1405)$ respectively
\begin{eqnarray}
\langle\eta'|\chi(0)|0 \rangle = -0.551+0.8208(1-{2c\over g})^{{1\over2}}\frac{m^2_{\eta'}}{m^2_\chi-m^2_{\eta'}}=0.3044
,\nonumber \\
\langle \eta(1405)|\chi(0)|0 \rangle = 0.8199+0.0.5724(1-{2c\over g})^{{1\over2}}\frac{m^2_G}{m^2_\chi-m^2_{G}}=-1.7788.
\end{eqnarray}
Eq. (19) show that the kinetic mixing
(the second terms of Eqs. (19)) plays an
essential role in those two matrix elements.
Inputting $\Gamma(J/\psi\rightarrow\gamma\eta')$, 
\(R=\frac{\Gamma(J/\psi\rightarrow\gamma \eta(1405))}{\Gamma(J/\psi\rightarrow\gamma \eta')}\)
is calculated. The ratio is very sensitive to the value of the mass of the c quark and
it increases with $m_c$ dramatically.
Few examples are presented
\begin{eqnarray}
m_c=1.22 \textrm{GeV},\;\;B(J/\psi\rightarrow\gamma \eta(1405))=3.67 (1 \pm 0.06)\times10^{-3},\nonumber \\
m_c=1.23 \textrm{GeV},\;\;B(J/\psi\rightarrow\gamma \eta(1405))=6.69 (1 \pm 0.06)\times10^{-3},\nonumber \\
m_c=1.24 \textrm{GeV},\;\;B(J/\psi\rightarrow\gamma \eta(1405))=1.13 (1 \pm 0.06)\times10^{-2}.
\end{eqnarray}
Large branching ratio for $J/\psi\rightarrow\gamma \eta(1405)$ is predicted for $m_c > 1.2 \textrm{GeV}$.

\section{$\eta(1405)\rightarrow\rho\pi\pi$ decay}
There are two subprocesses: (1) $\eta(1405)\rightarrow\rho\rho,\;\rho\rightarrow\pi\pi$, (2) $\eta(1405)\rightarrow\rho\pi\pi$
directly. Because $\chi\rightarrow\rho\rho$ vanishes only the $\eta_0\rightarrow\rho\rho$ contributes to (1). 
The amplitude of the subprocess (1) is derived as
\begin{equation}
T^{(1)} = -0.5724\frac{4\sqrt{6}}{\pi^2 g^3 f_\pi}
\frac{f_{\rho\pi\pi}}{q^2-m^2_\rho+i\sqrt{q^2}\Gamma(q^2)}\epsilon^{\mu\nu\alpha\beta}k_\mu e^\lambda_{\nu} k_{1\alpha}k_{2\beta},
\end{equation}
where 0.5724 is the coefficient of the $\eta_0$ component of $\eta(1405)$ (29), \(q = k_1 + k_2\), $\Gamma(q^2)$ is the
decay width of the $\rho$ meson. 
The subprocess (2) is the decay mode without intermediate resonance. The vertex of this process is similar to
$f_1\rightarrow\rho\pi\pi$ presented in Ref. [2] and it is found to be
\begin{equation}
{\cal L}_{\chi\rho\pi\pi} = \frac{2\sqrt{2}}{\sqrt{3}g\pi^2 f^2_\pi F}(1-{4c\over g})\epsilon_{ijk}
\epsilon^{\mu\nu\alpha\beta}\partial_\mu \chi
\partial_\nu\pi^i\partial_\alpha\pi^j\rho^k_\beta.
\end{equation}
The amplitude of the subprocess (2) is derived
\begin{equation}
T^{(2)} = 0.8199\frac{4\sqrt{2}}{\sqrt{3}g\pi^2}{1\over f^2_\pi}{1\over F}\epsilon^{\mu\nu\alpha\beta}p_\mu e^\lambda_\nu
k_{1\alpha}k_{2\beta},
\end{equation}
where 0.8199 is the coefficient of the $\chi$ component of $\eta(1405)$. Only the glueball component $\chi$ contributes
to $T^{(2)}$. Adding the two amplitudes together the amplitude of the process $\eta(1405)\rightarrow\rho^0\pi^+\pi^-$
is found and the decay width is computed
\begin{equation}
\Gamma(\eta(1405)\rightarrow\rho^0\pi^+\pi^-) = 0.92\; \textrm{MeV}.
\end{equation}
$T^{(1)}$ dominates the decay. The branching ratio of this channel is about $1.8\%$. 
The total branching ratio of $\eta(1405)\rightarrow\rho\pi\pi$ is $5.4\%$.

\section{$\eta(1405)\rightarrow a_0(980)\pi$ decay}
Two body decay,  $\eta(1405)\rightarrow a_0(980)\pi$, should be the major decay mode of $\eta(1405)$.
In the Lagrangian (1) the isovector scalar field $a_0(980)$ is not included and in order to study this
decay mode the $a_0(980)$ field must be introduced to the Lagrangian (1).
As mentioned in the section of introduction that a meson field is expressed as a quark operator in this theory. It is natural that
\begin{equation}
a_0(980) \sim \bar{\psi}\tau^i\psi a^i_0.
\end{equation}
The quantum numbers of $a_0(980)$ are \(J^{PC}=0^{++}\) and in the Lagrangian (1) there is already a term
\(-m\bar{\psi}u\psi\). The parameter m is originated in the quark condensate whose \(J^{PC}=0^{++}\) too.
It is proposed that the $a_0(980)$ field can be added
to the Lagrangian by modifying $-m\bar{\psi}u\psi$ to
\begin{equation}
-{1\over2}\bar{\psi}\{(m+\tau^i a_0^i)u+u(m+\tau^i a_0^i)\}\psi.
\end{equation}
At the tree level 
\begin{equation}
a^i_0=-{1\over m^2_{a_0}}\bar{\psi}\tau^i\psi.
\end{equation}
Therefore, Eq. (25) is revealed from this scheme.

The couplings between the mesons and the $a_0(980)$ can be derived. Using the vertex
\begin{equation}
{\cal L}=-\bar{\psi}\tau^i\psi a^i_0,
\end{equation}
the quark loop diagram of the S-matrix element $\langle a_0 |S| a_0 \rangle $ is calculated and
the kinetic term of the $a_0$ field is found. The $a_0$ field is normalized to be
\begin{equation}
a_0\rightarrow \sqrt{2\over 3}{1\over g}(1-{1\over 3\pi^2 g^2})^{-{1\over2}} a_0.
\end{equation}
The mass of the $a_0$ field is taken as a parameter.

The decay width of $\eta(1405)\rightarrow a_0\pi$ is calculated to the leading order in the momentum
expansion. Therefore, only
the $\eta_0\rightarrow a_0\pi$ channel is taken into account.
The vertices related to this channel are found 
\begin{eqnarray}
{\cal L}=-i\frac{2\sqrt{2}}{\sqrt{3}f_\pi}m\bar{\psi}\gamma_5\psi\eta_0
-i\frac{2m}{f_\pi}\bar{\psi}\tau^i\gamma_5\psi\pi^i\nonumber \\
-\sqrt{2\over3}{1\over g}(1-{1\over3\pi^2 g^2})^{-{1\over2}}\bar{\psi}\tau^i\psi a^i_0
-i{2m\over f_\pi}\sqrt{2\over3}{1\over g}(1-{1\over3\pi^2 g^2})^{-{1\over2}}\bar{\psi}I\gamma_5\psi a^i_0\pi^i,
\end{eqnarray}
where I is a $2\times 2$ unit matrix.
To the leading order in the momentum expansion the amplitude is found to be
\begin{equation}
T=-0.5724\frac{8\sqrt{2}}{\sqrt{3}f^2_\pi}{1\over g}(1-{1\over3\pi^2 g^2})^{-{1\over2}}
\{{1\over3}\langle0|\bar{\psi}\psi|0\rangle + 3m^3 g^2\},
\end{equation}
where the coefficient 0.5724 is the component of the $\eta_0$ of the $\eta(1405)$.
In the amplitude the quark condensate is obtained from the vertex $\bar{\psi}I\gamma_5\psi a^i_0\pi^i$ which is derived from
\begin{equation}
-{1\over2}i\bar{\psi}\gamma_5\{a_0 \pi+\pi a_0\}\psi.
\end{equation}
The vertices, $\bar{\psi}\tau^i\gamma_5\psi\pi^i$ and $\bar{\psi}\tau^i\psi a^i_0$
contribute to the term, $3m^3 g^2$. It is known that the quark
condensate is negative. Therefore, there is cancellation between the two terms of the amplitude (31). The cancellation
makes the decay width narrower. 
The decay width of $\eta(1405)\rightarrow a_0\pi$ is sensitive to the value of the quark condensate.
\[{1\over3}\langle0|\bar{\psi}\psi|0\rangle = -(0.24)^3\; \textrm{GeV}\]
is taken. 
\(m^2={f^2_\pi\over 6g^2(1-{2c\over g})^2}\) is determined in Ref. ~\cite{2}.
The total decay width of the three modes, $a^+_0\pi^-,\;a^-_0\pi^+,\;a^0_0\pi^0$ of $\eta(1405)\rightarrow a_0\pi$
is calculated to be
\begin{equation}
\Gamma(\eta(1405)\rightarrow a_0\pi)=44\; \textrm{MeV}.
\end{equation}
The branching ratio $B(\eta(1405)\rightarrow a_0\pi)=86 (1 \pm 0.07)\%$.
Therefore, $\eta(1405)\rightarrow a_0\pi$ is the major decay mode of $\eta(1405)$.

\section{$\eta(1405)\rightarrow K^*(890) K$ decay}
$\eta(1405)\rightarrow K^*(890) K$ is a possible
decay channel. The components of $\eta_0$ and $\chi$ are flavor singlets, only the component of $\eta_8$ contributes to the decay.
The proof can be found in Re. ~\cite{2}. 
The vertex obtained is
\begin{equation}
{\cal L}_{\eta(1405)K^*K} = a_3 cf_{ab8}\partial_\mu\eta_8 K^a_\mu K^b,
\end{equation}
where c is a constant proportional to \(a_3=0.00352\) which is from the component of the $\eta_8$ component of the
$\eta(1405)$.
oBViously, the contribution of this vertex to the decay $\eta(1405)\rightarrow K^* K$ is very small.
This theory predicts that the decay width of $\eta(1405)\rightarrow K^* K$ is very small.
A $0^{-+}$ resonance $\eta(1416)$ has been discovered in $\pi^- p\rightarrow K^+ K^-\pi^0 n$ at 18 $\textrm{GeV}$.
The parameters of this state are determined as 
\[M=1416 \pm 4 \pm 2\; \textrm{MeV},\;\;\Gamma=42 \pm 10 \pm 9 \;\textrm{MeV}.\]
These values are close to $\eta(1405)'s$. The ratio of the branching ratios
\begin{equation}
R=\frac{B(\eta(1416)\rightarrow K^* \bar{K} + c.c)}{B(\eta(1416)\rightarrow a_0\pi^0)}=0.084 \pm 0.024
\end{equation}
have been reported. The final state $a_0\pi$ has three states, therefore, the ratio should be divided by
3 and
\begin{equation}
R= 0.028 \pm 0.008.
\end{equation}
Assuming the $\eta(1416)$ is the $\eta(1405)$, $\Gamma(\eta(1405)\rightarrow K^* K)$ is narrower than
$\Gamma(\eta(1405)\rightarrow a_0\pi)$ by two orders of magnitude. This result supports the prediction made by this chiral
field theory.

\section{Summary}
Based on a phenomenologically successful chiral meson theory and the U(1) anomaly a chiral field theory of $0^{-+}$
glueball has been constructed. Systematic and quantitative study of the properties of the
candidate of the $0^{-+}$ glueball $\eta(1405)$ have been done by this theory.
The study of $\eta_8,\;\eta_0,\;\chi$ mixing shows that the
mass of $\eta(1405)$ fits the room of the pseudoscalar glueball well. The prediction of the small branching ratio
of $\eta(1405)\rightarrow2\gamma$ is consistent with the fact that $\eta(1405)$ has not been found in two photon collisions.
The theory predicts that $\eta(1405)\rightarrow a_0(980)\pi$ is the major decay mode of $\eta(1405)$.
A very small branching ratio of $\eta(1405)\rightarrow K^* K$ is predicted and the theory is consistent with the data.
The glueball component $\chi$ of the $\eta(1405)$ is the dominant contributor of
the $J/\psi\rightarrow\gamma\eta(1405)$ decay. The $K^* \bar{K}$ is the dominant decay mode of
the $\eta(1475)$.
Large $B(J/\psi\rightarrow\gamma\eta(1405))$ is via the kinetic mixing predicted. This is a very important channel to
identify the $\eta(1405)$ as a $0^{-+}$ glueball. As pointed in the section 7, there are two states $\eta(1405/1475)$
in the $\eta(1440)$ region. It is suggested
that measuring of the branching ratio of $J/\psi\rightarrow\gamma\eta(1405), \eta(1405)\rightarrow\delta\pi$
and $J/\psi\rightarrow\gamma\eta(1475), \eta(1475)\rightarrow K^*\bar{K}$ separately will be able to determine
$B(J/\psi\rightarrow\gamma\eta(1405))$. Then the decay modes of the $\eta(1405)$ can be measured. The comparison between the
experimental results and theoretical predictions should be able to determine whether the $\eta(1405)$ is, indeed, a $0^{-+}$
glueball.
The quark component $\eta_0$ of the $\eta(1405)$ is the dominant contributor of
the decay $\eta(1405)\rightarrow\gamma\gamma,\;\gamma V,\;\rho\pi\pi,\;a_0\pi$.
The glueball component $\chi$ of the $\eta(1405)$
is suppressed in these processes.
This chiral field theory can be
applied to study other possible candidates of the $0^{-+}$ glueball by input their masses into the theory to make
quantitative predictions.


\end{document}